# DWDM/OOC and large spectrum sources performance in broadband access network

AhmedD. KORA

Ecole Supérieure Multinationale des Télécommunications (ESMT), BP 10000
Dakar, Senegal
ahmed.kora@esmt.sn

## *ABSTRACT*

*This work presents a performance evaluation based on elaborated analytical expressions of error probability for broadband access networkin the case of a combined technique of dense wavelength division multiplexing (DWDM) and one dimensional optical orthogonal codes (1D-OOC). Optical sources with relatively large spectrum has been considered and simulated.Besides the Multiple Access Interference (MAI) at the receiver due to the access method which is optical code division multiple access (OCDMA), the emitted radiation of these sources in a dense WDM communication link introduces additional interference.Conventional correlation receiver (CCR) and parallel interference cancellation (PIC) receiverlimitations are discussed. This paper has investigated the kind of optical sources with large spectrum bandwidth which could be accepted for a targeted bit error rate (BER)and given number of users inbroadband access network supporting DWDM with optical orthogonal codes.*

## *KEYWORDS*

*OCDMA, OOC, DWDM, MAI, broadband, access network.*

## 1. INTRODUCTION

Thanks to information and communication technology advances, increasingly large amount of services and applications are offered to subscribers in telecommunication networks. More recently, it has been shown that context sources could provide valuable information about individuals and that can be used to drive adapted information to specific conditions and preference of each user [1]. This trend leads to a mismatch between wired networks and data traffic which becomes more and more important. It has become severe even with digital subscriber lines technologies denoted by (x DSL). Up to now, optical fiber is the media which presents the most capacity to enable this increasing data rate demand [2-10]. To overcome this problem firstly critical in long distance communication context, fiber optic has been deployedin transmission networks with technologies as synchronous optical network (SONET) and synchronous digital hierarchy (SDH). The limiting capacity of cooper to face the bandwidth demand has enabled the extension of fiber optic to access networks. But the slowly permeating of the fiber in access networks is related to the cost of optoelectronic components and civil engineering works of fiber deployment. In access network, fiber has been introduced first to overcome transport network limitation delimited by central office and the cabinet. This solution is known as FTTCab (Fiber To The Cab). The extension of fiber up to the curb corresponds to FTTC (Fiber To The Curb). When the fiber has reachedthe building, it has been called FTTB (Fiber To The Building). The FTTB technology assumed that cooper has been replaced by fiber from the central office to the users premises. An access network exclusively using optical fiber as media is depicted by FTTH/O (Fiber To The Home or Office). FTTH could be deployed based on different possible network topologies as point to point, loop … but the low cost FTTH solution is passive optical network (PON). Many generations of PON equipment could be found in the market : APON (Asynchronous transfer mode PON), BPON (Broadband PON), EPON (Ethernet PON), GPON (Gigabit PON), GEPON (Giga Ethernet PON), 10GPON, 10GEPON, …FTTx and PON have primarily used Time division multiple access (TDMA)  to get benefit from the bandwidth of the optical fiber in the access network. In order to improve bandwidth efficiency, wavelength division multiplexing (WDM) has been combined to TDMAin next generation PON (NG PON). Since the advent of 3G, optical code division multiple access (OCDMA) is also considered as a potential solution of PON evolution to NG PON. OCDMA is the CDMA sharing technology most applied to the context of optical access networks. This access method allows a large number of users depending on the code length to sharea common fiber in the access network.

A combined access technique as WDM/OCDMA has been assumed to achieve better bandwidth efficiency in NG PON.  One way of implementing WDM/OCDMA is to share an allocated transmitting wavelength among N users which generatedsignal could be differentiated by optical code spreading. Optical orthogonal codes performances have been investigated in many papers. In these works, the optical sources are supposed to be free of interference due to the width of their spectrum. The major drawback of using multiple sources with relatively wider spectrum width is the introduction of additional interference coming from the data transmitted by the other sources at the receiver bloc dedicated to a given transmitter source.

In this paper, two theoretical expressions of error probabilities for an opticalcommunication linkwith two kinds of receivers and relatively large spectrum of optical sourcesin a combined WDM/OOC system have been elaborated. By varying each parameter of the system, the effect of interference due to the proximity of central wavelengths is evaluated.The rest of this paper is organized as follows: Section 2 describes the considered WDM/OOC system architecture. In Section 3, theoretical error probability expressions have been elaborated. Simulations and results are provided in section 4. Then the paper ends with section 5 as Conclusion.

## 2.  SYSTEM DESCRIPTION

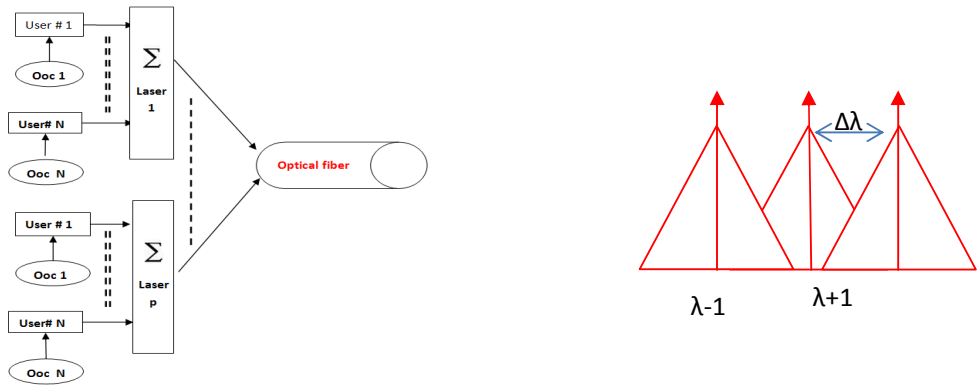

Fig. 1 OCDMA transmitting bloc descriptionFig. 2Optical sources spectrum

Fig. 1 depicts the transmitting bloc of the access network. A group of users allocated a maximum number *N*of optical orthogonal codes are supposed to transmit at the same wavelength. Their data are differentiated thanks to the spreading codes. In this system,the spectrum width of the different optical sources transmitting wavelength are supposed to overlap (Fig.2).

Each group of users sharing a given wavelength could belong to the same building. The aggregation of these signals from different buildings (Fig.2) up to central office will be a wavelength division multiplexing signal.It will be composed of several wavelengths in a single optical fiber in order to increase the capacity of the transmission link. On off keying modulation is considered in this work.

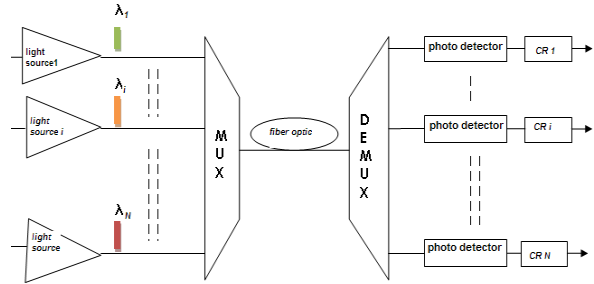

Fig. 3 Wavelength division multiplexing system description

At the receiver side, an appropriate common optical filter at each buildingor WDM demultiplexeris placed before photodetection. More bandwidth efficiency implies more wavelengths to be multiplexed and then fewer gaps between them. The consequence is additional interference due to channels with closer transmission wavelengths (Fig. 2).

Let consider an access network characterized by an incoherent and non-synchronous direct sequence OCDMA (DS-OCDMA) system. Each user data rate is given by D= $1/T_b$, where $T_b$ is the bit time duration. Before sending the users data over the optic fiber, each user data is spread by the destination allocated optical orthogonal code sequence called OOC [16], defined by four parameters (F, W, $h_a$, $h_c$). F is the length of the code; it can be calculated as the ratio of $T_b$ / $T_c$. This means that an OOC sequence is composed of F pulses or "chips". W is the code weight assuming that W of the F pulse are set to 1. The parameters $h_a$ and $h_c$ correspond to the auto-and cross-correlation constraints. It has been shown that the maximum number N of users depending on W and F in the case of $h_a=h_c=1$ is approximated in [17] by:

$$N = \frac{(F-1)}{W(W-1)} \qquad (1)$$

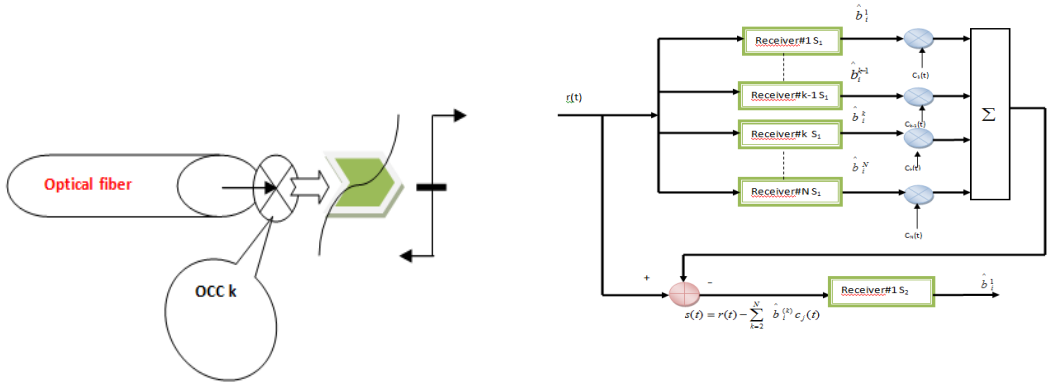

Fig. 4 Conventional Correlation Receiver description (CCR)      Fig.5 : A user Parallel Interference Cancellation receiver (PIC)

In the case of a same transmission wavelength down to the fiber, the received signal $r(t)$ is the aggregation of the different user's signal and the noise since they have been sent at the same wavelength.  After wavelength division demultiplexing, each receiver bloc recovers its dedicated data depending on the receiver system architecture. Two kinds of receivers are considered in this

paper.The conventional receiver (Fig. 4) and parallel interference cancellation (PIC) receiver (Fig.5).

The conventional correlation receiver (CCR) shown in Fig. 4 is known for its simplicity. Its dedicated data is recovered by multiplying the received signal by the signature corresponding to its attributed code sequence. Because of the lack of perfect orthogonality with OOC, the receiver output signal is subject to Multiuser Access Interference (MAI). Then it is integrated over the code length and compared to a decision threshold ''S'' in order to estimate the transmitted data. In the case of a synchronous system, conventional correlation receiver and no wavelength division multiplexing, the probability of error is given by (2) [18]:

$$P_e = \frac{1}{2} \sum_{i=s}^{N} C_i^{N-1} (\frac{W^2}{2F})^i (1 - \frac{W^2}{2F})^{N-1-i} \qquad (2)$$

The PIC receiver is more complex compared to CCR. It is composed of two main blocks. The first block is identical to the conventional receiver. In the second block, parallel interference cancellation (PIC) is performed.

Using OCDMA codes at different wavelengths increases the number of active users in the network. So, the degradation of the performance on a user link could not be only due to the effect of multiuser access interference (MAI) from the other users operating at the same wavelength but also from those of adjacent wavelength sources.

## 3. ANALYTICAL PROBABILITY ERROR EXPRESSIONS

The proposed study aims to be applied to broadband access network where components as Light-Emitting Diode with relatively large spectrum could be used. The simultaneous emission of these optical sources with central wavelength (..., $\lambda_{i+1}$, $\lambda_i$, $\lambda_{i-1}$,) in a DWDM (dense WDM) environment introduces additional interference at the receiver besides of multiple access interference (MAI). For simplicity,light sources with different wavelength but the same family of OOC sequence characterized by the quadruple (F, W, $h_a$, $h_c$) are considered. We have assumed that the system is chip based synchronous and similar family of OOC is used at each transmission wavelength. The power of a chip ''1'' is assimilated to unity.

### 3.1.Analytical probability error expressions with conventional correlation receiver

The conventional correlation receiver structure could be assimilated to two sub blocks where the contributions of the different sources are supposed separated by using the appropriate filter before applying the user's code to extract his data.The variable decision value related to a selected source at the receiver side using a conventional correlation receiver could be expressed as:

$$Z_i = Z_i^{(1)} + Z_i^{(2/1)} \qquad (3)$$

where

$$Z_i^{(1)} = W . b_i^{(1)} + \sum_{k=2}^{N} b_i^{(k)} \int_0^{T_b} c_k(t) c_1(t) dt \qquad (4)$$

is the decision variable component from the targeted source. In the case of only one disturbing source

$$Z_i^{(2/1)} = \sum_{k=1}^{N} b_i^{(k/2)} \int_0^{T_b} \alpha\, c_k(t)\, c_1(t)\, dt \qquad (5)$$

with $\alpha \in\ ]0,1[$, the detected disturbing source power level. Expression (5) represents the resulting component from other sources contribution.

The decoding rule could be expressed as:

$$\begin{cases} if\ Z_i \geq S \rightarrow \breve{b}_i^{(1)} = 1 \\ if\ Z_i < S \rightarrow \breve{b}_i^{(1)} = 0 \end{cases} \qquad (6)$$

With $\breve{b}_i^{(1)}$ the estimated bit and ''$S$'' the decision threshold of the receiver. The error probability is given by:

$$P_e = \frac{1}{2} prob(\breve{b}_i^{(1)} = 0/b_i^{(1)} = 1) + \frac{1}{2} prob(\breve{b}_i^{(1)} = 1/b_i^{(1)} = 0) \qquad (7)$$

$$= \frac{1}{2} prob(I_1 + \alpha * I_2 < S - W) + \frac{1}{2} prob(I_1 + \alpha * I_2 \geq S)$$

The first term in (7) is null since this event cannot happen. $I_1$ is an integer variable between 0 and $N$-1. Then

$$P_e = \frac{1}{2} \sum_{i=0}^{N-1} P(I_1 = i) * P(i + \alpha * I_2 \geq S) \qquad (8)$$

Let's $a \in N \setminus \{0,1\}$ ; $b \in N \setminus \{0\}$ with $\alpha = \frac{b}{a}$; $I_2$ is a random variable that is subject to binomial law which parameters are (9) and (10).

$$(N, \frac{W^2}{2L}) \qquad (9)$$

$$\Omega(I_2) = \{0, 1, 2, 3, \ldots\ldots\ldots\ldots, N\} \qquad (10)$$

So

$$P\left(I_2 \geq \frac{a}{b}(S-i)\right) = \sum_{I_2=\frac{a}{b}(S-i)}^{\infty} \sum_{h=0}^{N} C_N^h\, p^h (1-p)^{N-h} \delta(I_2 - h) \qquad (11)$$

$$P\left(I_2 \geq \frac{a}{b}(S-i)\right) = \sum_{h=\frac{a}{b}(S-i)}^{N} C_N^h\, p^h (1-p)^{N-h} \qquad (12)$$

The error probability expression yields to:

$$P_e = \frac{1}{2} \sum_{i=0}^{N-1} \sum_{h=\frac{a}{b}(s-i)}^{N} C_N^h C_{N-1}^i\, p^{i+h} (1-p)^{2N-(h+1+i)} \qquad (13)$$

With $= \frac{W^2}{2F}$ ; $N$: number of active user of the system; $F$: length of the code; $W$: weight of the code; $\alpha(b/a)$: wavelength multiplexing coefficient.

In the general case of ''$p$'' interfering optical sources with $\alpha_j$($j$ is an integer between 1 and $p$) the disturbing sources power level, we get

$$P_e = \frac{1}{2} \sum_{i=0}^{N-1} P(I_0 = i) * P\left(\left(i + \sum_{j=-q, j\neq 0}^{p} \alpha_j I_j\right) \geq S\right) \qquad (14)$$

189

In the case of $p$=5, the worst errorprobability expression derived from (14) could be rewritten as in [2]:

$$Pe = \frac{1}{2}\sum_{i=0}^{N-1}\sum_{u=0}^{N}\sum_{v=0}^{N}\sum_{w=0}^{N}\sum_{h=\frac{(S-i-\alpha_1 u+\alpha_{-1}v-\alpha_2 w)}{\alpha_{-2}}}^{N}C_{N-1}^{i} \tag{15}$$

$$* C_N^u C_N^v C_N^w C_N^h p^{i+u+v+w+h}\left(1-p\right)^{(5N-(i+u+v+w+h+1))}$$

## 3.2    Analytical probability error expression with PIC receiver

As described in section 2, the PIC receiver could be split into two sub blocks. The received signal after filtering is dispread by each OOC sequence in order to reproduce the interference of the undesired users. Supposingthe $k^{th}$ user as the desired one, the $N$-1 remaining undesired users are first detected thanks to the conventional correlation receiver (CCR) with a threshold $S$. Thereceiver builds and removes the expected MAI of the undesired users sharing the same WDM channel and then estimates the transmitted bit based on a CCR but with a second threshold $S_2$.The decision variable for a PIC receiver in the case of two optical sources could be expressed as:

$$Z_2^k = W.b_i^k + \sum_{\substack{j=1 \\ j\neq k}}^{N}(b_i^j - \hat{b}_i^j)\int_0^{T_b}c_j c_k dt + \sum_{x=1}^{N}\alpha d_i^x \int_0^{T_b}c_s c_k dt \tag{16}$$

Where $\hat{b}_i^j$ is the bit estimated with the first sub block as described in section 2.

The PIC receiver error probability expression yields to:

$$P_e^{PIC} = \left(\frac{1}{2}\right)^N \sum_{n_1=0}^{N-1}\sum_{n_2=W-S_2+1}^{N-n_1-1}C_{N-1}^{n_1}C_{N-1-n_1}^{n_2}P_l^{n_2}(1-P_l)^{N-n_1-n_2-1} \tag{17}$$

$$\times \left[\sum_{k_1=0}^{\left\langle\frac{S_2+n_2-W}{\alpha}\right\rangle}C_N^{k_1}R^{k_1}(1-R)^{N+k_1} + \sum_{\left\langle\frac{S_2+n_2}{\alpha}\right\rangle}^{N}C_N^{k_2}R^{k_2}(1-R)^{N+k_2}\right]$$

With$S_2$, the threshold of the PIC receiver in the second sub block; < x > the round function and

$$\begin{cases} P_l = R\sum_{j=0}^{n_1}\sum_{k_0=\left\langle\frac{S_1-j}{\alpha}\right\rangle}^{N}C_{n_1}^{j}C_N^{k_0}R^{j+k_0}(1-R)^{N+n_1-j-k_0} \\ R = \frac{W^2}{F} \end{cases} \tag{18}$$

# 4- SIMULATIONS AND RESULTS

This section is dedicated to numerical simulations results of the elaborated analytical expressions applied to broadband access network with a capacity greater than 40Gbps. The simulations show the impact of the spectrum width in different casesof WDM with DFB optical sources in 1550 nm transmission band. In these simulations a dense wavelength division multiplexing is considered and the gap between two adjacent wavelengths has been set to 0.8 nm. The spectrum bandwidth of the LASERS has been set to 2 nm. The case of OOCchannel without any WDM is also considered for comparison purpose.

New generation access networks might offer at least a capacity over 40Gbps. Depending on the length ''F'' of the designed OOC, the chip rate has to be greater than 40*F Gchips/s. In order to optimize users data rate we have considered a code length of 64 and we have set the code weight W=2. We get according to (1) an approximately maximum number of users similar to XPONswhich is 32.

The conventional correlation receiver (CCR) and parallel interference cancellation (PIC) receiver have been considered in various system configurations.

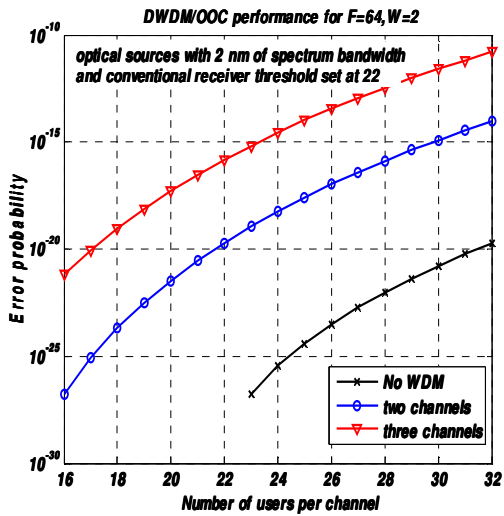

Figure 6. DWDM/ OOC channel performance with CCR threshold=22

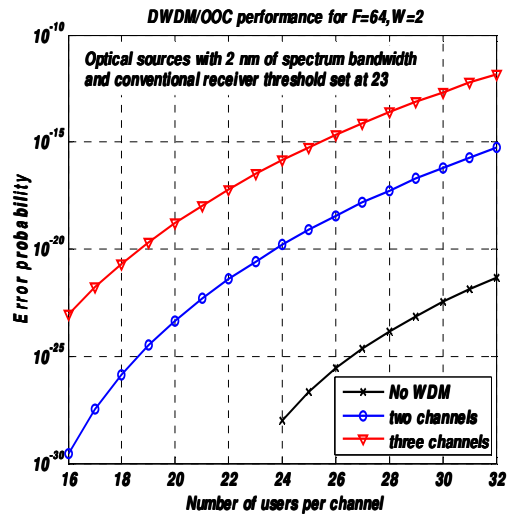

Figure 7. DWDM/ OOC channel performance with CCR threshold=23

All the simulation results (Fig.6 … Fig.12) confirm that increasing the number of user decrease the system performance. They also show that the best error probability performance in any case is the one without WDM.

All these simulations have pointed out that increasing the number of multiplexed wavelengths also degrades communication link quality.

Fig. 6 and Fig.7 exhibit DWDM/ OOC channel performance with CCR threshold valuefixed respectively at 22 and 23. Thesame value of code length and code weight is applied. From these two simulations results, it can be noticed that increasing the threshold level improves the communication channel performance. This threshold value also limits the maximum acceptable number of users at a targeted error probability.

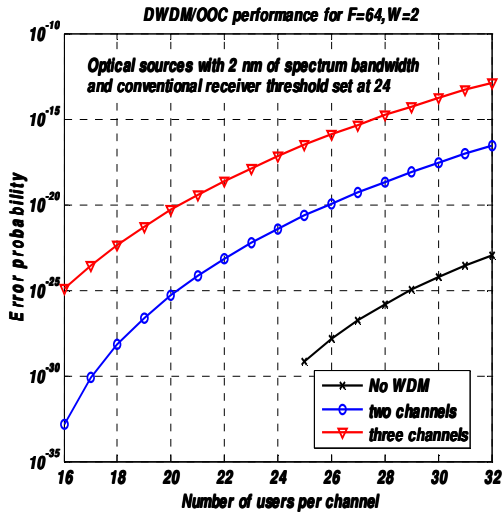

Figure 8. DWDM/ OOC channel performance with CCR threshold=24

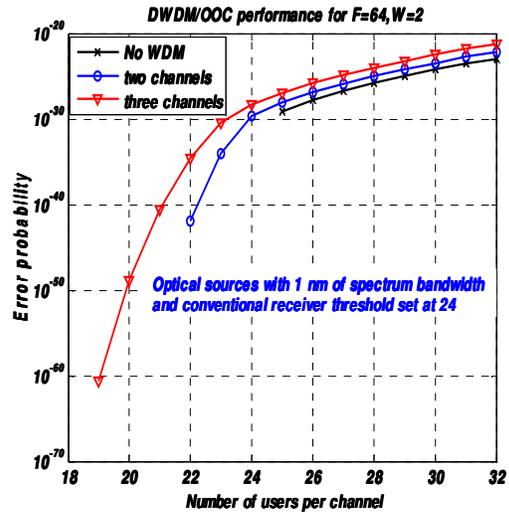

Figure 9. DWDM/ OOC channel performance with CCR threshold=24 and 1nm optical sources

Fig. 8 and 9 illustrate the effect of spectrum bandwidth. The CCR threshold in both simulations is 24 and the same system configuration is considered except that 2 nm optical source has been chosen in Fig.8 and 1nm optical source in the case of Fig.9.It can confirm that optical sources with less spectrum bandwidth at 3dB approximately equal to the gap between the adjacent multiplexing wavelengths attenuate drastically the impact of DWDM interference. It could be eliminated with designed optical sources.

In addition, Fig. 8 shows the minimum threshold value which could guarantee a maximum error probability of $10^{-12}$ with 32 users per channel for three multiplexed channels. The maximum supported number of user by the network is 96.

The impact of code length is depicted by Fig. 10 in comparison with Fig. 8. It shows that longer code improve the system link quality and capacity. The drawback is that it reduces the maximum data rate.

It is well known that increasing the code weight might decrease the performance loss since it introduces more possibility for two chips to be found at the same position.

Fig. 11 simulation results show performance loss for greater code weight at a same fixed threshold value in comparison with Fig 8. We can remark that the performance decreases since the maximum number of users set in (1) is exceeded. Figure 8 on the other hand presents better performance because an appropriate threshold is associated to the code weight. The considered large spectrum optical sources introduce enough interference.

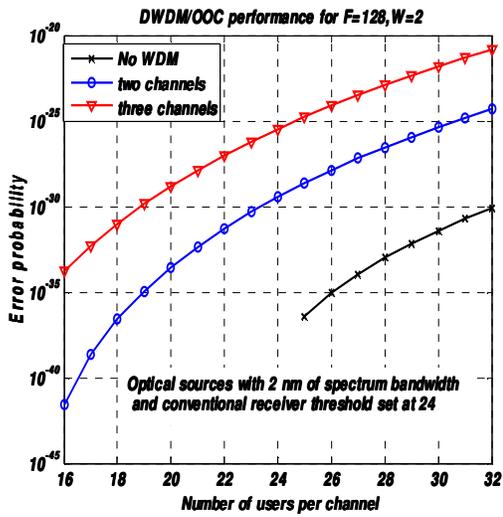

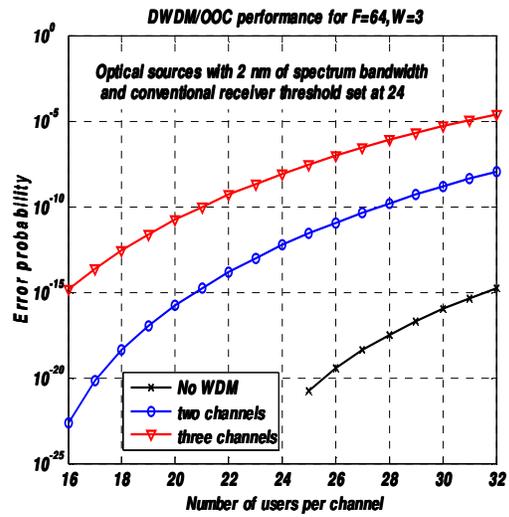

Figure 10. DWDM/ OOC channel performance with CCR threshold=24 and longer code length

Figure 11. DWDM/ OOC channel performance with CCR threshold=24 but greater code weight

Fig. 12 presents the performance comparison results between conventional correlation receiver and parallel interference (PIC) receiver in the case of DWDM and no WDM. The simulations results highlight clearly that PIC receiver outperforms CCR receiver. When the number of user increase, the performance of PIC receiver in case of WDM is closed to CCR receiver without no DWDM. But it is important to notice that a PIC receiver threshold might be carefully set in order to get good communication link performance.

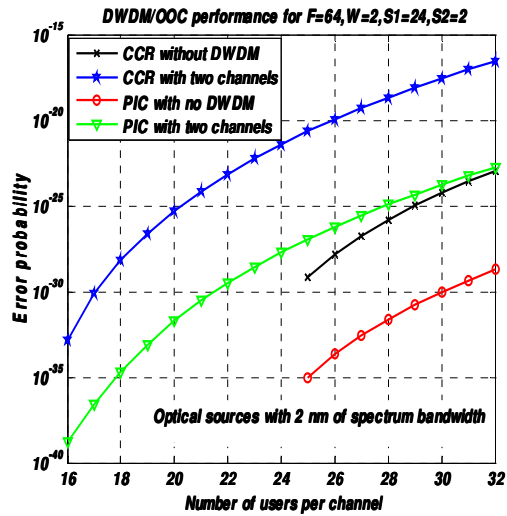

Figure12. DWDM/OOC channel performance comparison with different receivers

# 5.CONCLUSION

High data rate transmission in our everyday life has become crucial in access network. Broadband access network architecture with large spectrum optical sources using OOC family codes indense wavelength division multiplexing is considered in this paper. The error probability expressions have been presented and simulated. The performances of the system have been studied as function of an OOC system and WDM parameters. In order to lower the cost of transceivers, the main limitations of these WDM/OOC systems could be the spectrum width.Thanks to the various simulations,the optimal receiver thresholds of both CCR and PIC receivers at a probability error rate of $10^{-12}$based on the number of designed channels and number of users per channel could be estimated.It has been shown that 1 nm of spectrum width is critical to differentiate the performance of the system based on the number of multiplexed DWDM channels. Increasing the length of the user code limits the number of users in the network. Considering a minimum code weight of 2, the shortest code length in order to get 32 users has been set to 64. From this study, we have shown with our simulation results that it is possible to increase the number of user in a PON by using the combined WDM/OOC scheme at the cost of setting the right parameters to guarantee a good quality of the link.

# REFERENCES

[1] C. Baladron, J. M. Aguiar, B. Carro, L. Calavia, A. Cadenas and A. Sanchez-Esguevillas, (2012),''Framework for intelligent service adaptation to user's context in next generation networks'', IEEEComm. Mag., Vol. 50, No. 3, pp.18-25.

[2] Nadia Naowshin, A K M Arifuzzman, Mohammed Tarique, (2012) ''Demonstration and performance analysis of ROF based OFDM-PON system for next generation faber optic communication'' International Journal of Computer Networks & Communications (IJCNC), Vol. 4, No. 1, pp. 193-209.

[3] A. D. Kora, F. Diop, J.-P. Cances, S. Ouya, (2009), ''Optical Orthogonal Codes over Wavelength Division Multiplexing in passive Optical network communication system and its Performance analysis, 2nd IEEE International Conference on Adaptive Science & Technology'' (ICAST), pp. 40-44.

[4] S. Jindal, N. Gutpa, (2008),''Simulated Transmission Analysis 2D and 3D OOC for increasing Number of Potential Users'', *IEEE ICTON2008*, pp 302-305.

[5] S. V. Marie, M. D. Hahm, E. L. Titlebaum, (1995), ''Construction and Performance Analysis of a new family of Optical orthogonal Code for CDMA fiber Optic networks'', *IEEE Trans. On Comm*. Vol 43, No. 2/3/4, feb/March/April.

[6] Y. X. YANG, X. X. Niu, C. Q., (1997),''Counterexample of Truncated Costas Optical Orthogonal Codes'', *IEEE Trans. On Comm*. Vol 45, No.6.

[7]W. C. Kwong, and G.-C Yang, ''Design of Multilength Optical Orthogonal Codes for Optical CDMA Multimedia Networks'' *IEEE Trans. On Comm.,*Vol. 50,No. 8

[8] O. V. Sinkin, V. S. Grigoryan, C. R. Menyuk, (2007), ''Accurate Probabilistic treatment of Bit-Patter-Dependent Nonlinear Distorsions in BER calculations for WDM RZ Systems'', *J. Lightw. Technol.*, Vol. 25, No. 10, pp. 2959-2968.

[9] M. M. Karbassian and H. Ghafouri-Shiraz, (2007),''Performance Analysis of Heterodyne-detected Coherent Optical CDMA using a Novel Prime Code Family'', *J. Lightw. Technol.*, Vol. 25, N° 10, pp. 3028-3034,

[10] N.Saad,(2005), *Contribution à l'étude de l'application de la technique CDMA aux systèmes de transmission optique,* Thèse de Doctorat, Université de Limoges.

[11] B.Sarala, D.S.Venkateswarulu& B.N.Bhandari, (2012),''Overview of MC OCDMA PAPR reduction techniques'', International Journal of Distributed Parallel systems (IJDPS), Vol. 3, N°2, pp.193–206.

[12] MohamadDosaranianMoghadam, HamidrezaBakhshi&GholamrezaDadashzadeh, (2011), ''DS-CDMA cellular systems performance with Base Station Assignment, power control error and beamforming over multipath fading'',*International Journal of Computer Networks & Communications (IJCNC), Vol. 3, No. 1, pp. 185-202.*

[13] Davinder S Saini&Neeru Sharma, (2011),''Performance improvement in OVSF based CDMA networks using flexible assignment of data calls'',*International Journal of Computer Networks & Communications (IJCNC), Vol. 3, No. 6, pp. 197-212.*

[14] MohamadDosaranian-Moghadam, HamidrezaBakhshi, and GholamrezaDadashzadeh, (2010), ''Reverse Link Performance of DS-CDMA Cellular Systems through Closed-Loop Power Control and Beamforming in 2D Urban Environment'', International Journal of Computer Networks & Communications (IJCNC), Vol. 2, No. 6, pp. 136-153.

[15] Md. Sadek Ali, Md. Shariful Islam, Md. AlamgirHossain, Md. Khalid Hossain Jewel, (2011), ''BER analysis of multi-code multi carrier CDMA systems in multipath fading channel'', International Journal of Computer Networks & Communications (IJCNC), Vol. 3, No. 3, pp. 178-191.

[16] J.A.Salehi, (1989),''Emergingopticalcode-division multiple-access communicationsystems,'' *IEEE Netw. Mag.*, vol. 3, no. 2, pp.31–39.

[17] J. A. Salehi and C. A. Brackett, (1989),''Code-division multiple-access techniques in optical fiber networks-PartI:Fundamentalprinciples'',*IEEE Trans. Com.*, vol. 38, no. 8, pp. 824–833.

[18] C. Goursaud, A. Julien-Vergonjanne, C. Aupetit-Bertelomot, J.-P.Cances, J.-M. Dumas, (2006), ''DS-OCDMA Receivers based on Parallel Interference Cancellation and hard Limiters'', *IEEE Trans. on Com*., vol. 54, no. 9.

**Author**

Ahmed D. KORA is graduated in Physics Sciences in 1998 from "Faculté des Sciences Techniques" at "Universitéd'Abomey–Calavi", Bénin, where he got his Diplômed'EtudeApprofondie (DEA) in Material Sciences in 2000. In 2003, he received a Master "RéseauxTélécoms" degree from "EcoleSupérieure Multinational de Télécommunications" and the Ph.D. degree in telecommunication from the University of Limoges, France, in 2007. He is currently with the ''EcoleSuperieureMultinationale des Telecommunications''(ESMT). He is charge 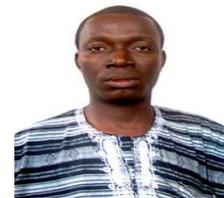 of coordinating the Graduate Program in Telecommunications named INGC. He has also trained a number of professionals from many international organizations and telecommunication operators on Fiber optic, Wimax, VSAT, Mobile networks, advanced network cabling… He is also in Instructor of Fiber Optic Association (FOA) and Global VSAT Forum (GVF) at ESMT. His research area covers optical communication and networks, channel estimation, iterative processing for MIMO communications, low cost microwave solutions and Wi-Fi propagation model.